\documentclass[12pt]{iopart}
\usepackage{graphicx,color}
\usepackage{bm}
\usepackage{braket}

\usepackage{hyperref}
\begin{document}

\title[Geometrical optimization of pumping power under adiabatic parameter driving]{Geometrical optimization of pumping power under adiabatic parameter driving}

\author{Masahiro Hasegawa \& Takeo Kato}

\address{Institute for Solid State Physics, The University of Tokyo, Kashiwa, Chiba 277-8581, Japan}
\ead{h.masahiro@issp.u-tokyo.ac.jp}
\vspace{10pt}
\begin{indented}
\item[]July 2019
\end{indented}

\begin{abstract}
Adiabatic pumping is a fundamental concept in the time-dependent transport of mesoscopic devices.
To maximize pumping performance, i.e., the amount of pumping per unit time, it is necessary to carefully manage the driving speed, which should be sufficiently less than the {\it limited speed}, an upper bound of the driving speed below which non-adiabatic effects are negligible.
In general, the amount of pumping increases as the contour of the driving parameter lengthens, however a long contour diminishes the pumping power because it requires more time per cycle under the {\it limited speed} constraint.
We consider this trade-off carefully and show that there should exist an optimized period and contour to maximize the power of adiabatic pumping.
We confirm this conclusion based on the results of charge pumping using a single-level quantum dot.
\end{abstract}

\section{Introduction}
\label{sec:intro}

Adiabatic pumping, a transport phenomenon induced by quasi-static periodic parameter driving, has attracted significant interest since it was first proposed by Thouless~\cite{Thouless1983May}.
In the field of mesoscopic physics, electron turnstile and pumping were studied both theoretically and experimentally by a number of researchers~\cite{Kouwenhoven1991Sep,Pothier1992Dec,Buttiker1993Jun,Büttiker1994Mar,Pretre1996Sep}.
For electronic transport via non-interacting quantum systems, the charge transference via adiabatic pumping is written in terms of the Berry curvature defined from the scattering matrices, which is known as Brouwer's formula~\cite{Brouwer1998Oct}.
This geometrical framework has been generalized to adiabatic charge pumping in various electron systems~\cite{Aleiner1998Aug,Wohlman2002Apr,Splettstoesser2005Dec,Splettstoesser2006Aug,Andergassen2010Jun,Pekola2013Oct,Hasegawa2017Jan,Hasegawa2018Mar}.
Recent developments in experimental techniques have also encouraged the study of adiabatic heat pumping via nanoscale devices~\cite{Ren2010Apr,Juergens2013Jun,Haupt2013Aug,Karimi2016Nov}.

Adiabatic pumping has also been studied as a basic concept of quasi-static operations in non-equilibrium steady state thermodynamics~\cite{Benenti2017Jun}.
Steady state entropy production induced by quasi-static operations has been formulated geometrically for general systems~\cite{Yuge2013Nov,Nakajima2017Dec}, and a non-equilibrium steady state entropy has been proposed for quantum dot~\cite{Esposito2015Feb} and harmonic oscillator systems~\cite{Ochoa2018Feb}.
The performance of quantum engines utilizing nanoscale devices has also been studied in phonon~\cite{Chamon2011Apr} and double quantum dot systems~\cite{Juergens2013Jun}.
The analysis of quantum engine performance has been recently extended to the non-adiabatic and non-equilibrium regions.
Notably, the trade-off relation between power and efficiency of engines has gained particular interest~\cite{Whitney2014Apr,Shiraishi2016Oct}.

Adiabatic pumping can be used in a number of applications, including electron and heat pumping, refrigeration, etc.
One important performance indicator of pumping devices is a power, i.e., a transfer rate of pumped quantities (heat or charge) per unit time.
In order to increase the power of adiabatic pumping, it is necessary to care about three factors: (a) driving speed, (b) shape of the pumping contour in parameter space, and (c) leakage current. 
We address each of these factors in turn as follows.

It is known that, at a given pumping contour within the parameter space, the amount of charge or heat transferred per cycle is constant and independent of the driving speed.
Accordingly, the adiabatic pumping power is proportional to the driving frequency.
There is, however, an upper limit on the driving speed determined by the adiabatic approximation condition.
In other words, the parameters should be driven quasi-statically;
if they are driven too quickly, non-adiabatic corrections become significant and the geometrical description of adiabatic pumping collapses.
Therefore, the pumping power of adiabatic pumping should be maximized under the constraint that the driving speed must be below a {\it limited speed}, the upper bound below which non-adiabatic corrections are negligible (for a precise discussion, see Appendix~\ref{apd:ad_app}).

\begin{figure}[tb]
\begin{center}
\includegraphics[width=10.0cm]{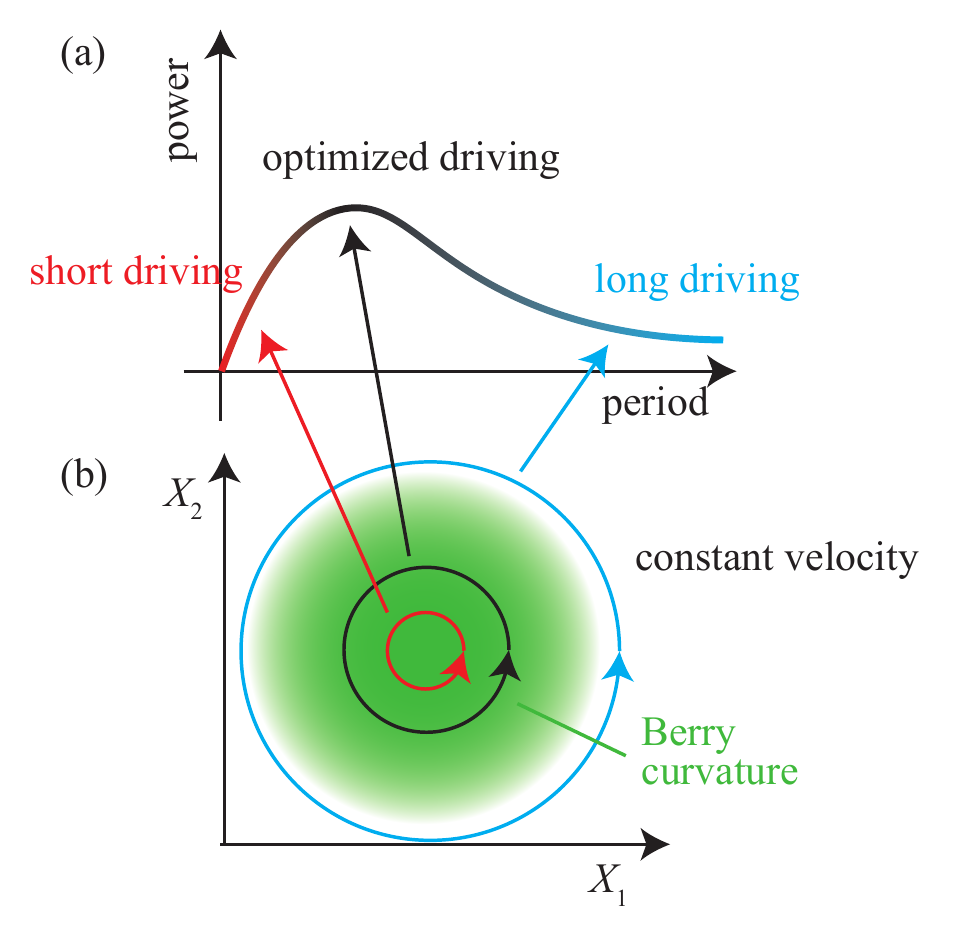}
\end{center}
\caption{\label{fig:abst}
(a) Schematic representation of the pumping power as a function of period per cycle.
(b) Examples of contours in the parameter space, where $X_1$ and $X_2$ denote driving parameters, and the density map plot (green) shows the Berry curvature as a function of $(X_1,X_2)$.
For short contours, the adiabatic pumping per cycle, which is characterized by the integral of the Berry curvature over the area enclosed by the contour, is so small that the pumping power is suppressed.
Pumping power is also reduced for long contours, because the pumping period becomes too large.
This suggests the existence of the optimized driving period with the optimized contour at which the power is maximized.
}
\end{figure}

The second factor, the shape of the contour loop in the parameter space, directly determines the pumping power via Brouwer's formula~\cite{Brouwer1998Oct}, in which the amount of charge or heat transferred per cycle is expressed by a surface integral of the Berry curvature over the area enclosed by the contour.
We note that the pumping power can be written in terms of the Berry curvature even for interacting electron systems~\cite{Splettstoesser2005Dec,Hasegawa2017Jan,Hasegawa2018Mar} (see Appendix~\ref{apd:ad_app}).
The amount of pumping per cycle increases when the contour encloses a wide area on which the Berry curvature takes a large value. 
Although it would seem that the maximum power of pumping is achieved when the area enclosed by the contour is as large as possible, this strategy fails because the driving speed must be below the {\it limited speed} discussed in the previous paragraph.
Although the amount of pumping per cycle increases with contour size, the period is proportional to the length of the contour under the condition that the driving speed is set as a constant value below the {\it limited speed}.
Therefore, the pumping power can be diminished when the contour length is too large.
This discussion indicates that there is an optimal contour and period which maximize the pumping power (see figure~\ref{fig:abst}).
This is the main subject of this paper.

The third factor, the leakage current, also affects the optimal contour.
Charge (heat) pumping devices may be operated against a finite bias for chemical potentials or temperatures.
In such biased systems, the leakage charge (heat) current flows in the direction opposite to that of pumping.
The leakage current is given by a time integral over the contour, and depends on both the cycle period and the amplitude of leakage current.
Therefore, to suppress the leakage current under a fixed speed of driving as possible, the optimal contour should be either shortened or avoid the parameter region on which leakage current is large.

The purpose and approach of this paper are summarized as follows.
We consider optimization of the power in adiabatic pumping under the constraint that the driving speed remains below the {\it limited speed}, which guarantees adiabaticity in system dynamics induced by parameter driving.
Increasing the contour length increases the amount of pumping per cycle, which contradicts the goal of increasing the pumping rate, the number of cycles per unit time.
Large contours can also diminish the pumping power because they increase the leakage current flowing against the pumping direction.
Taking these factors into account, we consider the optimization problem for general types of pumping.
We also demonstrate the existence of the optimal contour by looking at a simple example of electron pumping using a non-interacting single-level quantum dot.
Our discussion is restricted to the trade-off relation between the amount of adiabatic pumping and the pumping period, which should be distinguished from the trade-off relation discussed in reference~\cite{Shiraishi2016Oct}.

This paper is organized as follows:
In section~\ref{sec:formalism}, we introduce a general framework for optimizing the pumping power.
In section~\ref{sec:application}, we discuss the adiabatic charge pumping via a single-level quantum dot system using approximate calculation.
In section~\ref{sec:summary}, we summarize our results.

\section{Formalism \label{sec:formalism}}

In this section, we introduce a general framework for optimizing pumping power.
For simplicity, we consider two-parameter pumping although it is straightforward to extend this framework into pumping by more than two parameters.
Throughout this paper, we employ the unit of $\hbar = 1$.

\subsection{Adiabatic approximation}

We consider pumping induced by two-parameter driving described by the dimensionless parameter vector, $\bm{X}(t) = ( X_1(t),X_2(t) )$.
The parameter driving is periodic in time, i.e., $\bm{X}(t+{\cal T}) = \bm{X}(t)$, where ${\cal T}$ is the period of the pumping cycle. 
We employ the adiabatic approximation~\cite{Brouwer1998Oct,Splettstoesser2005Dec,Hasegawa2017Jan,Hasegawa2018Mar}, in which it is assumed that the parameter driving is sufficiently slow relative to the characteristic time scales of the system.
In this approximation, the amount of charge or heat transferred per cycle, $Q$, is given by the sum of two contributions:
\begin{eqnarray}
	Q &\simeq Q^{\mathrm{st.}} + Q^{\mathrm{ad.}}, \label{eqn:add_app_pumping} \\
	Q^{\mathrm{st.}} &=  \int_{0}^{\cal T} \! dt \, j [ \bm{X}(t) ], \label{eqn:steady_contribution} \\
	Q^{\mathrm{ad.}} &= \int_{A} dX_1 dX_2 \, \Pi (\bm{X}), \label{eqn:Brouwers_formula}
\end{eqnarray}
where $Q^{\mathrm{st.}}$ is the total amount of a steady state leakage current per cycle, $Q^{\mathrm{ad.}}$ is the contribution of adiabatic pumping, $j[\bm{X}(t)]$ is the steady state leakage current, and $\Pi(\bm{X})$ is the Berry curvature.
Equation~(\ref{eqn:Brouwers_formula}) is known as Brouwer's formula~\cite{Brouwer1998Oct}, in which the transferred heat or charge is described as the surface integral of $\Pi(\bm{X})$ over the two-dimensional area $A$ enclosed by the driving contour $C = \{ \bm{X}(t)\, |\, 0 \leq t < {\cal T} \}$ (for a detailed derivation, see Appendix~\ref{apd:ad_app}).
Whereas the steady state leakage depends on the driving speed of ${\bm X}(t)$, the contribution of adiabatic pumping is independent of it.

\subsection{Optimization functional}

To parametrize the contour, $C = \{ \bm{X}(s)\, |\, 0 \leq s < 1 \}$, we introduce a dimensionless parameter $s$ and rewrite the time integral as
\begin{eqnarray}
	\int_{t_i}^{t_f} dt \to \int_0^1 ds \ \tau(s) \label{eqn:time_to_normalize} ,
\end{eqnarray}
where $\tau(s)=dt/ds$ is a function of $s$.
We note that $\tau(s)ds$ denotes a dwell time on the contour interval $[s,s+ds]$, and depends on the speed of the parameter driving.
For simplicity, we assume, as described in the introduction, that the driving velocity $v=|d{\bm X}/dt|$ is constant. 
The amount of transferred charge or heat per cycle, $Q$, is then given by a functional of the contour $C$ as
\begin{eqnarray}
	Q[C] = \int_{0}^{1} ds \ l(s) U[\bm{X}(s)] + \int_A dX_1 dX_2 \ \Pi (\bm{X}) . \label{eqn:cost_func}
\end{eqnarray}
where $l(s)=v\tau(s)$, and $U[\bm{X}(s)]$ is the normalized leakage current defined as
\begin{eqnarray}
	U[\bm{X}(s)] = v^{-1} j[\bm{X}(s)] .
\end{eqnarray}

To optimize the adiabatic pumping under the fixed driving period ${\cal T}$, we introduce the following functional based on the Lagrange multiplier,
\begin{eqnarray}
	Z[C,{\cal T}] &= Q[C] + \lambda \left[ v^{-1}  \int_{0}^{1} ds \ l(s) - {\cal T} \right], \label{eqn:cost_func_LM}
\end{eqnarray}
where $\lambda$ is the Lagrange multiplier.
Using the optimized contour $C_{\rm opt}$ that maximizes $Z[C,{\cal T}]$, the total pumping power $P$ is defined as
\begin{eqnarray}
	P({\cal T}) = \frac{Z[C_{\mathrm{opt}},{\cal T}]}{{\cal T}} .
	\label{eq:PumpingPower}
\end{eqnarray}
Our main purpose is then to clarify whether there is a `best' period ${\cal T}_{\rm max}$ for maximizing $P({\cal T})$.

\section{Single-level quantum dot system \label{sec:application}}

\begin{figure}
\begin{center}
\includegraphics[width=10.0cm]{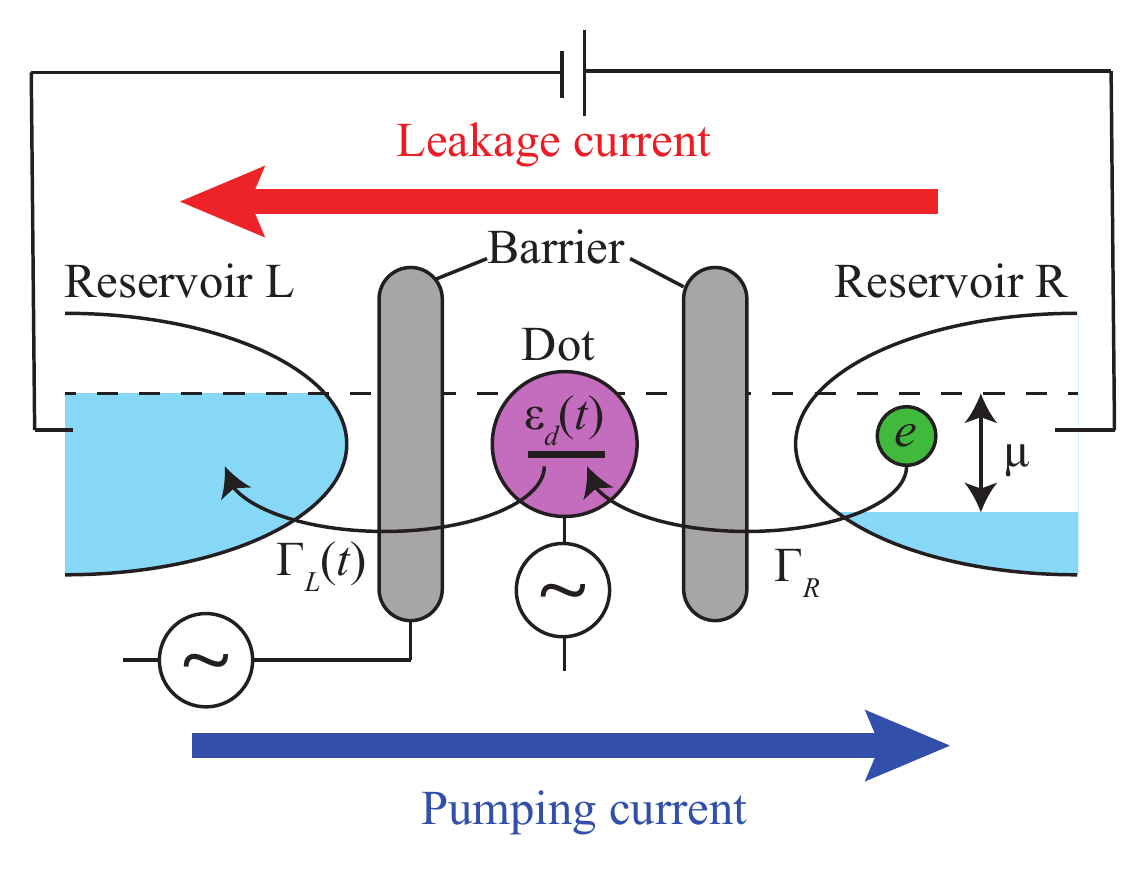}
\end{center}
\caption{\label{fig:model}
Schematic of the single-level quantum dot system.
A quantum dot and two electron reservoirs, reservoir $L$ and $R$, are separated by potential barriers and coupled to each other via electron tunneling.
The energy level in the quantum dot, $\epsilon_d (t)$, and the coupling between the dot and the reservoir $L$, $\Gamma_L (t)$, are taken as time-dependent driving parameters.
The bias between reservoirs is expressed as a chemical potential difference $\mu$.
The leakage current induced by the bias flows from the reservoir $R$ to $L$, while the charge is pumped in the opposite direction (note that the direction of the electron transfer is opposite).
The sign of the current is defined as positive for the pumping current (the blue arrow) and negative for the leakage (the red arrow).
}
\end{figure}

In this section, we consider the pumping performance under a specific setup---electron pumping via a single-level quantum dot (see figure~\ref{fig:model})---and demonstrate that there indeed exists an optimizing period that maximizes the pumping power.
As the optimal period is obtained by a simple mechanism (see also section~\ref{sec:intro}), we can expect that the result obtained in this section also holds for general systems.

\subsection{Model}

The Hamiltonian, which describes charge transport via a single-level quantum dot, is given as
\begin{eqnarray}
	H(t) = H_d(t) + H_{t}(t) + \sum_{r=L,R} H_r ,
\end{eqnarray}
where
\begin{eqnarray}
	H_d(t) &= \epsilon_d(t) d^{\dagger} d , \\
	H_{t}(t) &=  \sum_k [ v_{L}(t) c_{Lk}^{\dagger} d + v_{R} c_{Rk}^{\dagger} d  + h.c. ],\\
	H_r &= \sum_{k} \epsilon_{k} c^{\dagger}_{rk} c_{rk} .
\end{eqnarray}
Here $d$ ($d^{\dagger}$) denotes the annihilation (creation) operator of an electron in the quantum dot and $c_{rk}$ ($c_{rk}^{\dagger}$) denotes the annihilation (creation) operator of electrons in reservoir $r = L,R$ with wave number $k$ and energy $\epsilon_{k}$.
In our model, we assume that the energy level of the quantum dot, $\epsilon_d (t)$, and one of the dot-reservoir couplings, $v_L(t)$, are time-dependent driving parameters, whereas the other dot-reservoir coupling, $v_R$, is taken to be constant.
We further assume that both of the reservoirs are kept at zero temperature and have their Fermi distribution functions given as
\begin{eqnarray}
	f_L(\epsilon) = \Theta (\epsilon - \mu/2) , \quad f_R(\epsilon) = \Theta (\epsilon + \mu/2) ,
\end{eqnarray}
where $\Theta(\epsilon)$ is the Heaviside step function, and the average of the Fermi energies of the two reservoirs is set to zero.
In the wide-band limit, the linewidth of the energy level in the quantum dot is given as $\Gamma_L(t) + \Gamma_R$, where
\begin{eqnarray}
	\Gamma_L(t) &= \pi \rho |v_{L}(t)|^2 , \\
	\Gamma_R &= \pi \rho |v_{R}|^2 \equiv \Gamma .
\end{eqnarray}
Here $\rho$ is the reservoir density of states at the Fermi energy.
We employ $\Gamma$ as a unit of the energy in the following calculations.

We consider charge pumping from reservoir $L$ to $R$ against the chemical potential bias $\mu (>0)$.
The sign of the charge current is defined as positive when it flows from $L$ to $R$ (which corresponds to electrons transferring from $R$ to $L$).
Under the present setup, the pumping current is positive and the leakage current is negative (see figure~\ref{fig:model}).

\subsection{Leakage current and Berry curvature}

We introduce the following dimensionless driving parameters:
\begin{eqnarray}
X_1(t) &= x(t) = \epsilon_d(t) / \Gamma, \\
X_2(t) &= y(t) = \Gamma_L(t) / \Gamma.
\end{eqnarray}
Under these definitions, the driving velocity $v=|d{\bm X}/dt|$ has a dimension of the energy, and should be taken as the {\it limited speed}.
The {\it limited speed} should be determined rigorously by the adiabatic condition (equation~ (\ref{eqn:adaibatic_condition})). 
However, such rigorous  {\it limited speed} is not practical and roughly estimated one is enough to show the trade-off relation.
In the following calculation, we estimate it roughly as $v=0.1\Gamma$, assuming that $\Gamma$ is a typical energy scale of the present system.

%Under these definitions, the driving velocity $v=|d{\bm X}/dt|$ has a dimension of the energy, and should be taken as the {\it limited speed}, i.e., it should be sufficiently smaller than the characteristic energy scale of the system, $\Gamma$
%\footnote{
%We should note that the present estimate of the {\it limited speed} is rough. 
%The {\it limited speed} should be determined rigorously by the adiabatic condition (equation~ (\ref{eqn:adaibatic_condition})). 
%In a simple case, the coefficients, $J_{n}^{(2,1)}$ and $J_{n}^{(2,1)}$, are typically proportional to the inverse of the characteristic energy scale of the system. 
%However, the coefficients cannot be estimated by this simple consideration, since they may strongly depend on the parameter vector when the driving amplitude is large (e.g., see reference~\cite{Moskalets2002Nov,Splettstoesser2008Nov}).
%This indicates that careful consideration is needed to determine the {\it limited speed}.
%Rigorous consideration of non-adiabatic effect, which is required especially for pumping with a long contour, gives more strict condition for the {\it limited speed} in general models.
%}.
%In the following calculation, we set it as $v=0.1\Gamma$.

By the Meir-Wingreen formula~\cite{Meir1992Apr}, the normalized leakage current is calculated as
\begin{eqnarray}
	U(x,y) = \frac{e}{2\pi} \frac{2 \Gamma y}{v} \int_{-\omega_0}^{\omega_0} \! d\omega \ \mathcal{A}(\omega) , \label{eqn:pot_gen}
\end{eqnarray}
where $\omega_0 = \mu \Gamma^{-1} / 2$, $e\ (< 0)$ is the elementary charge, and $\mathcal{A}(\omega)$ is the normalized spectral function:
\begin{eqnarray}
	\mathcal{A}(\omega) = \frac{1}{(\omega-x)^2 + (1+y)^2/4} .
\end{eqnarray}
Applying the adiabatic approximation in the Keldysh Green's function approach~\cite{Splettstoesser2005Dec,Hasegawa2017Jan,Hasegawa2018Mar}, the Berry curvature is calculated as
\begin{eqnarray}
	\Pi(x,y) &=  - \frac{e}{2\pi} \Bigl[ y \mathcal{A}^2(\omega_0) + \frac{(1-y)}{2} \mathcal{A}^2(-\omega_0) \nonumber \\
	&\hspace{1.5cm} - 2 \int_{-\omega_0}^{\omega_0} d\omega \ y (\omega-x)  \mathcal{A}^3(\omega) \Bigr]  . \label{eqn:Pi_gen}
\end{eqnarray}

\subsection{Optimization by an elliptical contour}

From the functional $Z[C,{\cal T}]$ given in equation~(\ref{eqn:cost_func_LM}), we can derive the differential equations to be satisfied for the optimized contour $C_{\rm opt}({\cal T})$ under the constraint of a fixed ${\cal T}$ (see Appendix~\ref{apx:stationary_eqn}).
The optimized contour $C_{\rm opt}({\cal T})$ is then obtained by numerically solving these equations, and the pumping power $P(T)=Z[C_{\rm opt},{\cal T}]/{\cal T}$ is obtained as a function of the period ${\cal T}$.
For the present purpose of demonstrating the existence of the contour with the optimal pumping performance, it is sufficient to approximate the contour as a simple shape.
In this section, we restrict the contour shape to an ellipse in the parameter space $(x,y)$, and optimize parameters of the ellipse, namely, the position of the center, the lengths of the semi-major and semi-miner axes, and the angle of the semi-major axis, to maximize $Z[C,{\cal T}]$ (see Appendix~\ref{apx:rep_of_E_and_HE}).

\begin{figure}
\begin{center}
\includegraphics[width=10.0cm]{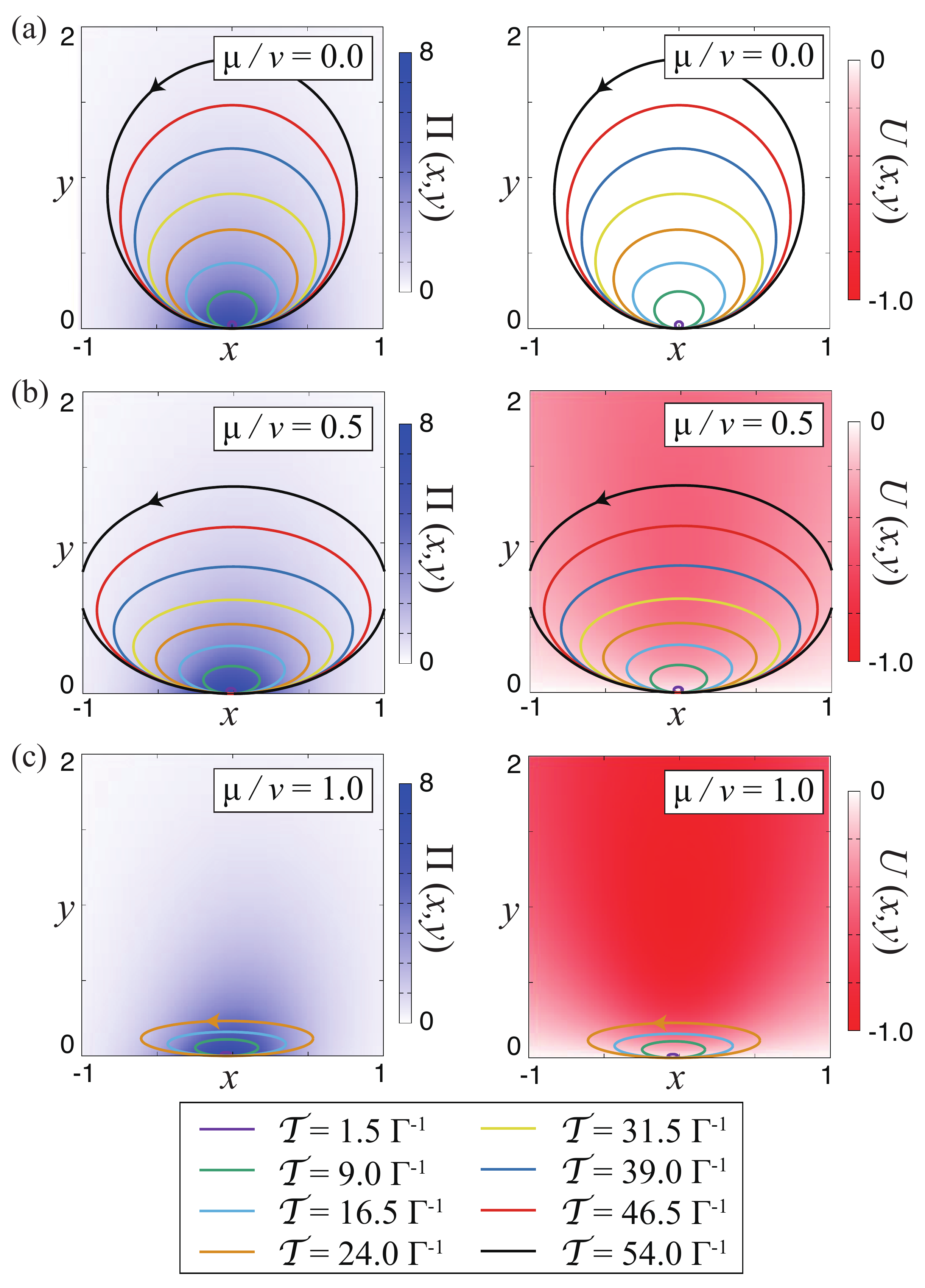}
\end{center}
\caption{\label{fig:contourC}
Optimized elliptical contours for eight periods from ${\cal T} = 1.5 \Gamma^{-1}$ to ${\cal T} = 54.0 \Gamma^{-1}$.
The velocity $v$ is set to $0.1 \Gamma$ and the chemical potential bias is set to (a) $\mu / v = 0.0$ , (b) $\mu / v = 0.5$, and (c) $\mu / v = 1.0$.
The left and right panels show the Berry curvature $\Pi(x,y)$ and the normalized leakage current $U(x,y)$, respectively, as a density plot.
Both functions are represented in units of $|e|/2\pi$.
For $\mu / v = 1.0$, charge pumping is possible for only four short-period contours, because the leakage current is always dominant for $\mathcal{T} > 28.5 \Gamma^{-1}$.
}
\end{figure}

Figure~\ref{fig:contourC} shows the optimized elliptical contours obtained under several pumping periods for three chemical potential biases: (a) $\mu / v = 0$ , (b) $\mu / v = 0.5$ , and (c) $\mu / v = 1.0$.
The color density plots in the left and right panels show the Berry curvature $\Pi(x,y)$ and the normalized leakage current $U(x,y)$, respectively.
The optimized contours (solid lines) are the same in the left and right panels.

In the unbiased case (figure~\ref{fig:contourC}~(a)), the contours reflect only the profile of the Berry curvature in the parameter space because there is no leakage current.
All of the optimized contours touch the origin because the Berry curvature has its maximum at the origin.
As the period increases, the contours gradually expand along the $y$-direction.
The Berry curvature is evaluated as $(1+4x^2)^{-2}\Pi(0,0)$ for $y=0$ and  $(1+y)^{-3}\Pi(0,0)$ for $x=0$, respectively.
Since $\Pi(x,y)$ decays faster in the $x$-direction, it is advantageous for charge pumping to extend the contour in the $y$-direction as the driving period ${\cal T}$ increases.

In the biased cases (figure~\ref{fig:contourC}~(b) and (c)), the leakage current becomes finite and is given as a monotonically increasing function of $y$ for fixed $x$.
As a result, the optimized contour shrinks toward the line $y=0$ to avoid the loss of pumped charge due to the leakage as possible.
The degree of the shrink becomes larger as the bias increases (figure~\ref{fig:contourC}~(b) and (c)).
We should note that only four contours are presented in figure~\ref{fig:contourC}~(c), which indicates that charge pumping is impossible for long-period contours where the leakage dominates the charge pumping.

\begin{figure}
\begin{center}
\includegraphics[width=10.0cm]{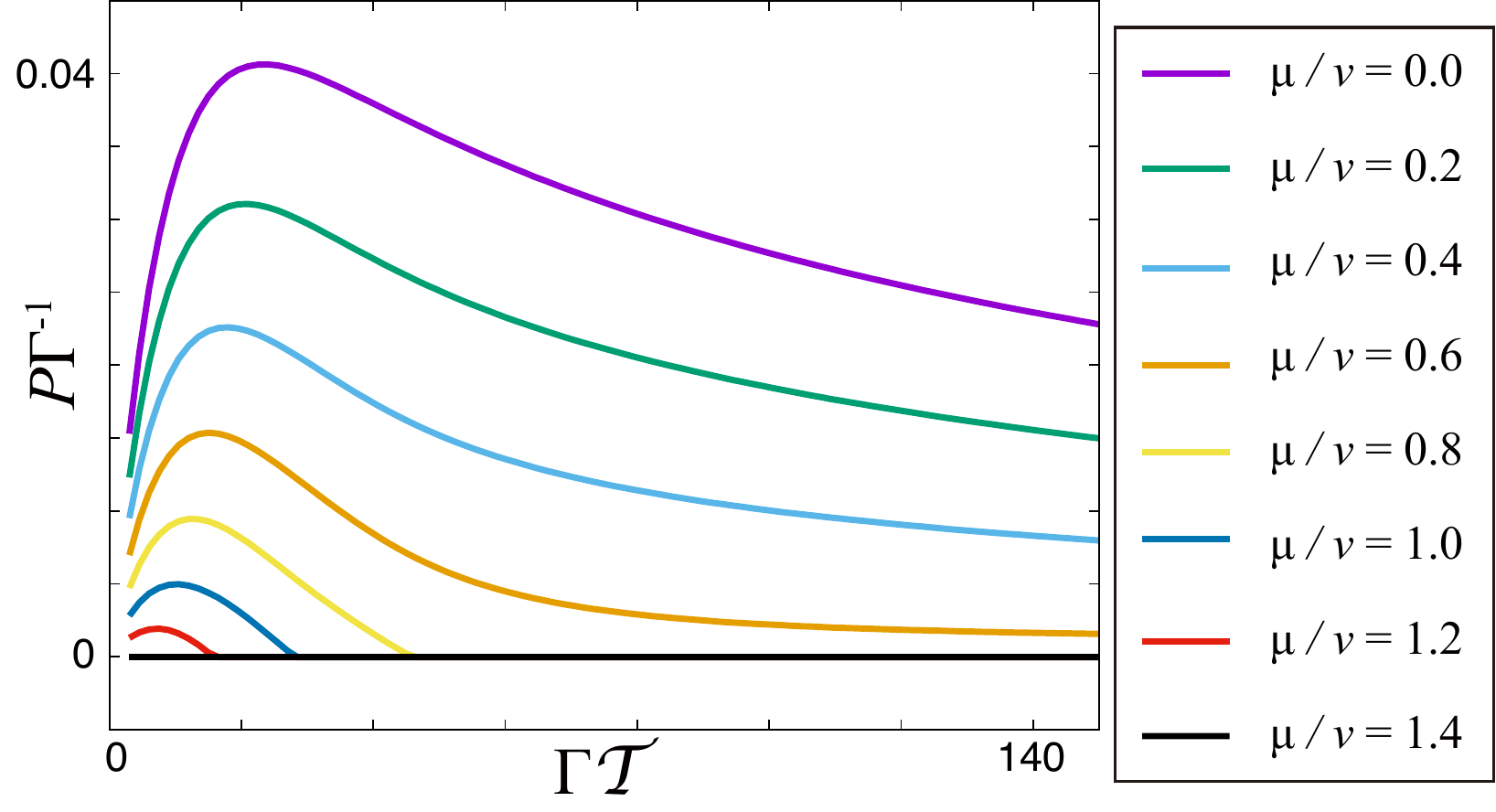}
\end{center}
\caption{\label{fig:effC}
Pumping power $P$ for elliptical contours as a function of the period ${\cal T}$ in units of $|e|/2\pi$.
The chemical potential bias is set to eight values from $\mu / v = 0.0$ to 1.4.
The velocity is set to $v = 0.1 \Gamma$.
}
\end{figure}

Figure~\ref{fig:effC} shows the pumping power $P({\cal T})=Z[C_{\rm opt},{\cal T}]/{\cal T}$ as a function of ${\cal T}$ for several chemical potential biases.
For the unbiased case ($\mu=0$), the pumping power has a maximum at the optimal value of the period, ${\cal T}_{\rm opt}$.
The existence of this optimal period is explained as follows (see also section~\ref{sec:intro}).
The pumped charge ``per cycle" increases monotonically as the period ${\cal T}$ becomes longer because the area enclosed by the contour, over which the Berry curvature is integrated, is enlarged.
Therefore, as the period increases, the pumping power increases for ${\cal T}<{\cal T}_{\rm opt}$.
For longer-period pumping (${\cal T}>{\cal T}_{\rm opt}$), however, the pumping power is degraded as the denominator of the pumping power in equation~(\ref{eq:PumpingPower}) becomes dominant.
It should be noted that the optimal period ${\cal T}_{\rm opt}$ is on the order of the inverse of the {\it limited speed} $v$ (here, we set it as $v=0.1\Gamma$).
As the chemical potential bias $\mu$ increases, the pumping power is suppressed, and the optimal period ${\cal T}_{\rm opt}$ decreases for $\mu / v < 1.2$.
The charge pumping against the bias becomes impossible in the whole region for $\mu / v = 1.4$.

\subsection{Optimization by a half-ellipse contour}

\begin{figure}
\begin{center}
\includegraphics[width=10.0cm]{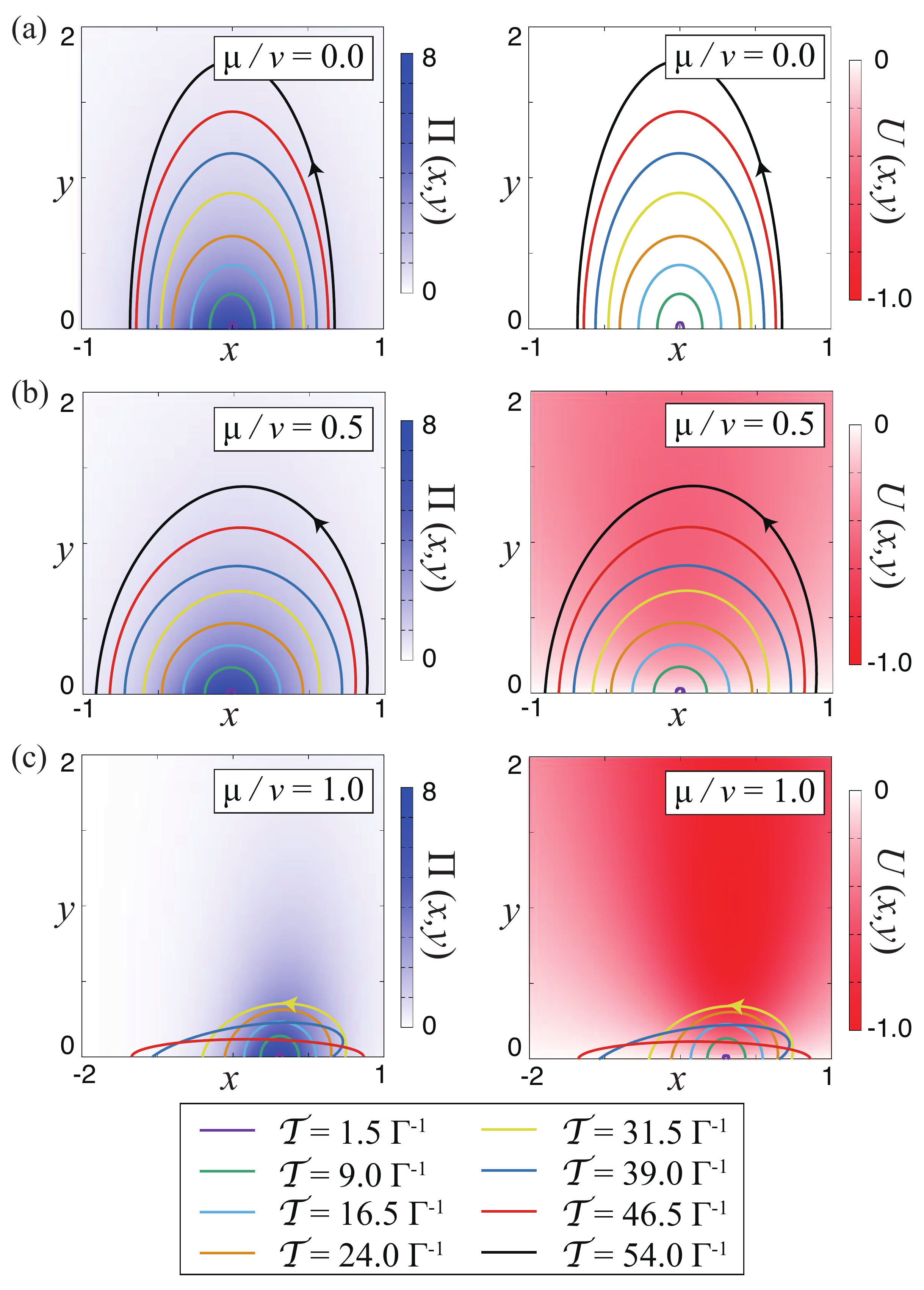}
\end{center}
\caption{\label{fig:contourEHC}
Optimized half-elliptical contours for eight periods.
The parameter setup is the same as that in figure~\ref{fig:contourC}.
The left and right panel show the Berry curvature and the normalized leakage current, respectively, as a density plot in units of $|e|/2\pi$.
For $\mu / v = 1.0$, charge pumping is possible for only seven short-period contours, because the leakage current is always dominant for ${\cal T} > 54.0\Gamma^{-1}$.
}
\end{figure}

The existence of the optimal period for charge pumping is the main result of this paper.
This result should not depend on the choice of the contour shape.
To demonstrate this, we consider a half-elliptical contour (for details, see Appendix~\ref{apx:rep_of_E_and_HE}), and compare the results with those obtained for an elliptical contour.

Figure~\ref{fig:contourEHC}~(a)-(c) show optimized contours for several pumping periods, with density plots of the Berry curvature and leakage current given in the left- and right-hand panels, respectively.
The parameters are the same as those used in the elliptical contour cases, and the overall features are similar to the results obtained in these cases.
For the unbiased case (figure~\ref{fig:contourEHC}~(a)), the contours enclose a region near the origin, in which the Berry curvature has a maximum value.
As the chemical potential bias $\mu$ increases (figure~\ref{fig:contourEHC}~(b) and (c)), the contours shrink toward the line $y=0$ to avoid the region where the leakage current is large.
The centers and inclination angle of the half-elliptical contours changes more clearly than the elliptical contours, reflecting small changes in the Berry curvature and leakage current.

\begin{figure}
\begin{center}
\includegraphics[width=10.0cm]{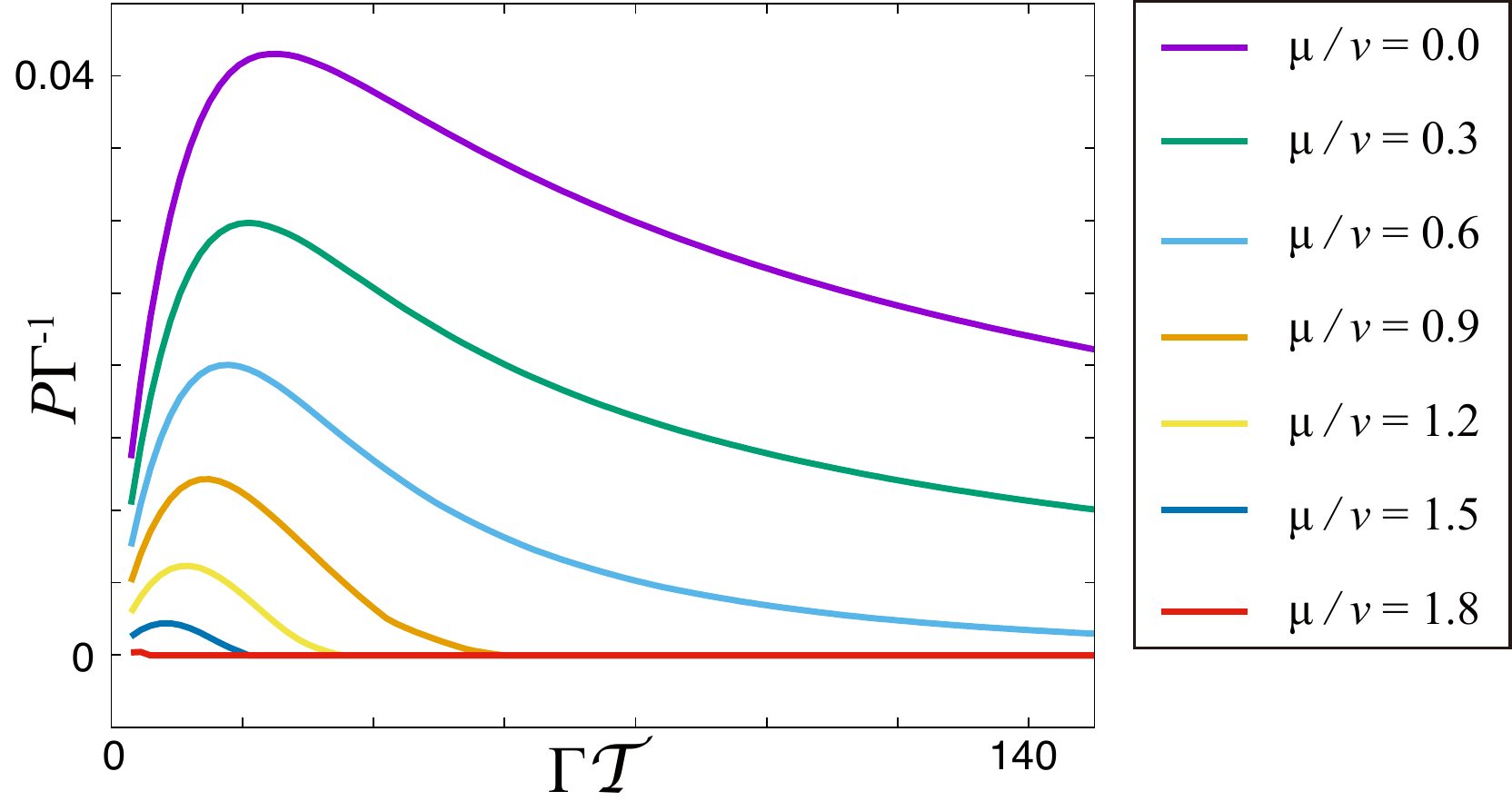}
\end{center}
\caption{\label{fig:effEHC}
Pumping power $P$ for half-elliptical contours as a function of the period ${\cal T}$ in units of $|e|/2\pi$.
The chemical potential bias is set to seven values from $\mu / v = 0.0$ to 1.8.
The velocity is set to $v = 0.1 \Gamma$.
}
\end{figure}

Figure~\ref{fig:effEHC} shows the pumping power of the half-elliptical contours for several biases.
While the features remain in qualitative agreement with those obtained in the elliptical case, the pumping power against the bias is improved and pumping is possible up to $\mu / v \sim 1.8$, a larger value than in the elliptical case.
The maximum value of pumping power, $P_{\rm opt}\simeq 0.041$, is also slightly larger than that in the elliptical case ($P_{\rm opt}\simeq 0.040$).

\section{Summary \label{sec:summary}}

In this paper, we studied the optimization of the performance of adiabatic pumping.
We formulated the optimized contour under a fixed pumping period assuming that the velocity of the parameter driving is constant.
We then derived the optimal contour using an actual system involving charge pumping via a single-level quantum dot by approximating contour shapes as ellipses and half-ellipses.
This analysis revealed that pumping power achieves maxima at a specific optimal driving period in the unbiased case, and that, under small levels of chemical potential biases, it is possible to pump charges against the bias though the leakage current opposes the charge pumping.
Because the optimal contour and period are determined by a simple mechanism, we expect that these features derived for a specific example are applicable to general transport systems.

The adiabatic condition for the driving velocity is roughly estimated in this paper.
It remains as a future problem to derive the detailed condition for the adiabatic pumping, and to improve the formulation of optimization.

\clearpage

\appendix

\section{Adiabatic approximation \label{apd:ad_app}}

In this appendix, we briefly derive  equations~(\ref{eqn:add_app_pumping})-(\ref{eqn:Brouwers_formula}) for general systems including interacting electron systems
(see Refs.~\cite{Splettstoesser2005Dec,Hasegawa2017Jan,Hasegawa2018Mar} for a detailed derivation).
For simplicity, we assume that the time-dependent parameter vector $\bm{X}(t)$ is driven in the sinusoidal manner:
\begin{eqnarray}
    X_n(t) = X_{n,0} + \delta X_n \sin (\omega t +\phi_n).
\end{eqnarray}
Here $X_{n,0}$, $\delta X_n$ and $\phi_n$ are the center, amplitude, and phase of the sinusoidal parameter driving, respectively, and $\omega$ is the pumping frequency.
Although this assumption appears to oversimplify the problem, it captures the essence of the adiabatic approximation.
Assuming that the parameter vector $\bm{X}(t)$ is driven slowly, we can consider adiabatic expansion for the time-dependent current $J(t)$ as
\begin{eqnarray}
    J(t) &= J^{(0)} + \sum_n J^{(1)}_{n} \dot{X}_n(t) + \sum_n J^{(2,1)}_{n} \ddot{X}_n(t) \nonumber \\
    &\hspace{0.5cm} + \sum_{n,m} J^{(2,2)}_{nm} \dot{X}_n(t) \dot{X}_m(t) + \cdots .
\end{eqnarray}
In the adiabatic approximation, the leading and the next leading term, $J^{(0)}$ and $J^{(1)}_n$, are taken into account while the remaining terms are neglected.
To justify this approximation, the pumping frequency and amplitude must satisfy the following condition:
\begin{eqnarray}
    &J_{n}^{(2,1)} \omega^2 \delta X_{n}, \ J_{n}^{(2,2)} \omega^{2} \delta X_{n} \delta X_{m} \  \ll  J_{n'}^{(1)} \omega \delta X_{n'} , \label{eqn:adaibatic_condition}
\end{eqnarray}
for arbitrary $m$, $n$ and $n'$.
This condition implies a trade-off between the amplitude and the frequency of pumping; the pumping frequency should be reduced if the driving amplitude increases, and vice versa.
This relation indicates that the product of amplitude and frequency, $\omega \, \delta {\bm X} \simeq \dot{\bm X}$, has an upper bound, which we call the {\it limited speed}.
For the dimensionless parameter vector ${\bm X}$, this {\it limited speed} is mostly taken as much smaller than the characteristic energy scale of the system.

We next examine the leading and next leading terms in detail.
The leading contribution corresponds to the steady state contribution,
\begin{eqnarray}
	J^{(0)}(t) = j[\bm{X}(t)] = \left. J(t) \right|_{\mathrm{st.}}. \label{eqn:ad_app_st}
\end{eqnarray}
Here $\left. O \right|_{\mathrm{st.}}$ denotes the steady-state average of the observable $O$ under a fixed parameter vector $\bm{X}(t)$.
The next leading contribution corresponds to the contribution of the adiabatic pumping:
\begin{eqnarray}
	J^{(1)}_n = \pi_n [\bm{X}(t)] = \left. \frac{\partial J(t)}{\partial \dot{X}_n(t)} \right|_{\mathrm{st.}} , \label{eqn:ad_app_ad}
\end{eqnarray}
where $\pi_n [\bm{X}(t)]$ is called the Berry connection.
The amount of charge or heat transferred per cycle is described as the sum of the steady and adiabatic part:
\begin{eqnarray}
Q &= Q^{\rm st.} + Q^{\rm ad.}, \\
Q^{\rm st.} &= \int_{0}^{\mathcal{T}} dt \ J^{(0)}(t) \label{eqn:charge_transfer_gen} ,\\
Q^{\rm ad.} & = \int_{0}^{\mathcal{T}} dt \ J^{(1)}(t) = \oint_C d{\bm X} \cdot {\bm \pi},
\end{eqnarray}
where $\mathcal{T}$ is the pumping period, which equals to $2\pi \omega^{-1}$ in this case, and $C$ is a closed contour in the parameter space.
For the two-parameter driving ${\bm X}=(X_1,X_2)$, we can rewrite the formula for $Q^{\rm ad.}$ using the Stokes theorem as
\begin{eqnarray} 
Q^{\rm ad.} = \int_{A} dX_1 dX_2 \, \Pi({\bm X}),
\end{eqnarray}
where $\Pi({\bm X}) = \partial_{X_1} \pi_2({\bm X}) - \partial_{X_2} \pi_1({\bm X})$ is the Berry curvature, and $A$ is the area enclosed by the contour $C$.
Thus, we obtain equations~(\ref{eqn:add_app_pumping})-(\ref{eqn:Brouwers_formula}).

As an example, we consider charge pumping via a non-interacting electron system using the Brouwer's formula~\cite{Brouwer1998Oct}.
The amount of charge pumped from the reservoir $L$ to $R$ is then written in terms of the scattering matrix as
\begin{eqnarray}
	\pi_m[\bm{X}(t)] &= \frac{e}{\pi} \ \sum_{r=L,R} \mathrm{Im} \Bigl[ \frac{\partial S_{L r}}{\partial X_{m}} S_{L r}^{\dagger}  \Bigr]  .  \label{eqn:app_ber_con_brouwer}
\end{eqnarray}
Here $S_{r_1 r_2}$ denotes a scattering matrix element from reservoir $r_1$ to $r_2$.
Although the scattering theory is not applicable to interacting electron systems, Keldysh Green's function method can be used to generalize this formula even to interacting systems, and it gives the same form as equations~(\ref{eqn:add_app_pumping})-(\ref{eqn:Brouwers_formula}) (for details, see  Refs.~\cite{Splettstoesser2005Dec,Hasegawa2018Mar}).
It is also worth noting that it is possible to obtain the same geometrical formalism for an incoherent transport system by using Master equation approach~\cite{Juergens2013Jun}.

\section{Stationary equation \label{apx:stationary_eqn}}
\begin{figure}
\begin{center}
\includegraphics[width=10.0cm]{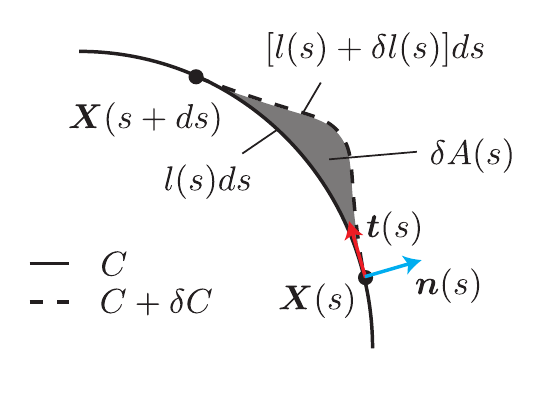}
\end{center}
\caption{\label{fig:variation}
Schematic of contour variation.
By adding a variation $\delta \bm{X}(s)$ within the infinitesimal section $[s,s+ds]$, the contour $C$ (solid line) is changed to the contour $C + \delta C$ (dotted line).
Under this variation, the length of the contour increases from $l(s) ds$ to $[l(s)+\delta l(s)]ds$.
The area of the contour also increases by $\delta A(s)$.
The red and blue arrows denote the tangent vector $\bm{t}(s)$ and the normal vector ${\bm n}(s)$ of the contour $C$, respectively.
}
\end{figure}

In this appendix, we derive the stationary equation to optimize  $Z[C,\mathcal{T}]$. 
Let us consider small variation of contours, $C\rightarrow C+\delta C$ (see figure \ref{fig:variation}).
This is described by variation of the parameter vector, $\bm{X}(s) \to \bm{X}(s) + \delta \bm{X}(s) $.
Here we can describe the variation $\delta \bm{X}(s)$ as a linear combination of the tangent vector $\bm{t}(s)$ and the normal vector $\bm{n}(s)$ of contour $C$ as
\begin{eqnarray}
	\delta \bm{X}(s) = \delta t(s) \bm{t}(s) + \delta n(s) \bm{n}(s) .
\end{eqnarray}
We choose an orientation of ${\bm n}(s)$ such that the area $A$ enclosed by $C$ increase when $\delta n(s)$ is positive.
The difference in $Z[C,\mathcal{T}]$ induced by this variation is evaluated in terms of three contributions as follows:
\begin{eqnarray}
	\delta Z[C,\mathcal{T}] &= Z[C+\delta C,\mathcal{T}] - Z[C,\mathcal{T}] \nonumber \\
	&\simeq \int_{0}^{1} ds \Bigl\{ ( U[\bm{X}(s)] + \lambda v^{-1} ) \delta l (s)  \nonumber \\
	&\hspace{1.5cm} + l(s) \delta U[\bm{X}(s)] + \Pi (\bm{X}) \delta A(s) \Bigr\} .
\end{eqnarray}
Here $\delta A \simeq  \delta n(s) l(s)$ is the variation in the area, and $\delta l (s)$ is the variation in the line element,
\begin{eqnarray}
	\delta l(s) &= \left| \frac{d ( \bm{X}(s) + \delta {\bm X}(s))}{ds} \right| - \left| \frac{d \bm{X}(s)}{ds} \right|  \nonumber \\
	&\simeq  \dot{\delta t }(s) + (\bm{t}(s) \cdot \dot{\bm{n}}(s) ) \delta n(s)  .
\end{eqnarray}
Hereafter we use the notation, $\dot{f} = df /ds $.
The variation of the normalized leakage current is then written as 
\begin{eqnarray}
	\delta U[\bm{X}(s)]  &= U[\bm{X}(s)+\delta X(s)] - U[\bm{X}(s)]  \nonumber \\
	&\simeq \partial_t U[\bm{X}(s)] \delta t(s) + \partial_{n} U[\bm{X}(s)] \delta n(s) .
\end{eqnarray}
Here $\partial_{t/n}$ is the directional derivative along the tangent/normal directions, respectively.
The stationary condition, $\delta Z [C{,\mathcal{T}}] = 0$, leads to the equation
\begin{eqnarray}
	&(\bm{t}(s) \cdot \dot{\bm{n}}(s) ) ( U[\bm{X}(s)] + \lambda v^{-1}) \nonumber \\
	&\hspace{1.0cm} + l(s) \partial_{n} U[\bm{X}(s)] + l(s) \Pi [\bm{X}(s)] = 0 . \label{eqn:stationary_equation}
\end{eqnarray}
By solving this equation, we can obtain the contour $C$ that maximize the amount of pumping under a fixed period,  $Z[C,{\cal T}]$.

\section{Representation of ellipse and half-ellipse \label{apx:rep_of_E_and_HE}}

\begin{figure}[tbh]
\begin{center}
\includegraphics[width=10.0cm]{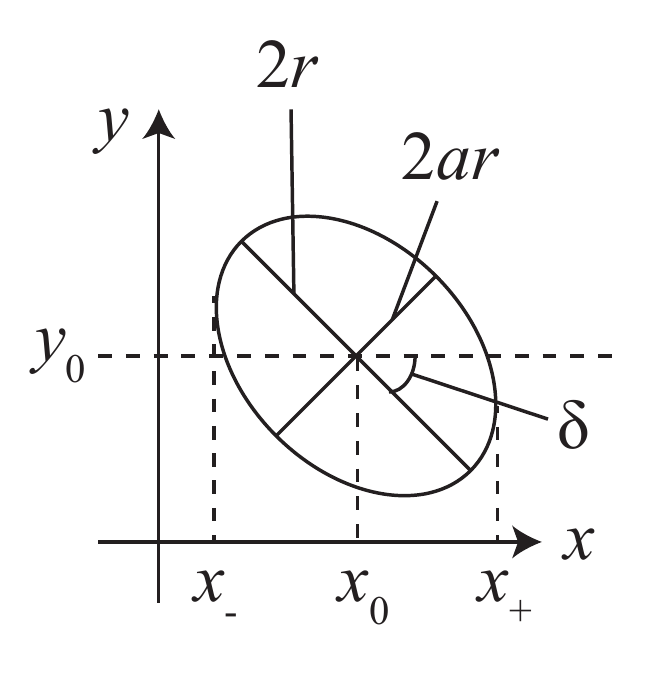}
\end{center}
\caption{\label{fig:ellipse}
Parametrization of  an elliptical contour.
}
\end{figure}

We can describe the elliptical contour by the following parametric representation as following (see figure~\ref{fig:ellipse}):
\begin{eqnarray}
	& x(s) = r(s) \cos 2\pi s + x_0 , \\
	& y(s) = r(s) \sin 2\pi s  + y_0,
\end{eqnarray}
where
\begin{eqnarray}
	r(s) = \frac{r}{\sqrt{\cos^2 (2\pi s + \delta) + a^2 \sin^2 (2\pi s + \delta) }} .
\end{eqnarray}
The parameters, $x_0$, $y_0$, and $\delta$, vary under the condition that all the points on the contour are located in the upper plane ($y\ge 0$).

We describe the half-ellipse by
setting $y_0$ to zero, and by restricting the parameter $s$ to the range $0 \leq s \leq 0.5$:
\begin{eqnarray}
	& x(s) = r(s) \cos 2\pi s + x_0 , \\
	& y(s) = r(s) \sin 2\pi s .
\end{eqnarray}
The remaining contour is a straight line described by
\begin{eqnarray}
	& x(s) = (x_+ - x_-)(s-0.5) + x_- , \\
	& y(s) = 0 .
\end{eqnarray}
for $0.5 \leq s \leq 1$, where $x_+$ and $x_-$ are the intersection points between the ellipse and the horizontal line passing through its center (see figure~\ref{fig:ellipse}).

\ack

The authors acknowledge M. Moskalets for helpful comments on the discussion of adiabatic approximation and T. Tamaya for useful comments on the manuscript.
M.H. acknowledges financial support provided by the Advanced Leading Graduate Course for Photon Science.
T.K. was supported by JSPS Grants-in-Aid for Scientific Research (No. JP24540316 and JP26220711).

\section*{References}
\bibliography{references}

\end{document}